\documentclass[epj]{svjour}
\usepackage{color}
\usepackage{graphics}
\usepackage{amssymb}
\newcommand{\sss}{\scriptscriptstyle}
\newcommand{\sst}{\scriptstyle}
\begin{document}  
\title{The role of hyperon resonances in $p \left( \gamma, K^+ \right)
\Lambda$ processes}
\author{Stijn Janssen\thanks{\email{stijn.janssen@rug.ac.be}}, Jan
Ryckebusch, Wim Van Nespen, Dimitri Debruyne and Tim Van Cauteren}  
\institute{Department of Subatomic and Radiation Physics \\
Ghent University, Proeftuinstraat 86, B-9000 Gent, Belgium} 
\date{\today}
\abstract{We discuss the role of hyperon resonances in the $u$-channel
when modeling $p \left( \gamma, K^+ \right) \Lambda$ processes in an
effective Lagrangian approach.  Without the introduction of hyperon
resonances, one is forced to use soft hadronic form
factors with a cutoff mass which is at best two times the kaon mass.
After inclusion of the hyperon resonances in the $u$-channel, we obtain 
a fair description of the data with a cutoff mass of the order of 1.8
GeV.}
\PACS{ {13.60.Le}{} 	
	{14.20.Jn}{} 	
	{14.20.Gk}{} 	
}
\authorrunning{Stijn Janssen {\em et al.}}

\maketitle
\section{Introduction}
\label{sec:introduction}
Photo-production of strangeness on the nucleon is a potentially
powerful tool for studying hadrons at the ``con\-sti\-tu\-ent-quark'' scale
of $\sim$ 1~GeV \cite{Geiger}. It is hoped that through
comparing model 
calculations with sufficiently large sets of $p \left( \gamma, K^+
\right) \Lambda$ data \cite{Tran}, our understanding of the excitation
spectrum 
and the structure of the nucleon will be deepened. Regarding our
knowledge about the excitation spectrum of the nucleon,
strangeness production provides a study domain for resonances
which remain undiscovered in $\pi$ photo-production or $\pi N
\rightarrow \pi N$ scattering reactions. 
At variance with the $(\gamma,\pi)$ reaction, even at threshold the
invariant energy of the $p \left( \gamma, K^+ \right) \Lambda$
reaction exceeds the mass of several hadron resonances. 
Accordingly, in modeling $p \left( \gamma, K^+ \right) \Lambda$
processes in terms 
of hadronic degrees-of-freedom, a considerable part of the  excitation
spectrum of 
the nucleon can in principle participate in the reaction mechanism.
For some of these resonances, the existence and branching into the
strange channels is well established. For others, no convincing
empirical evidence in support of their existence could as yet be
produced.  When calculating $p \left( \gamma, K^+ \right) \Lambda$
observables within the framework of effective field theories, one
frequently employs various combinations of hadron resonances.  This
procedure may be perceived as cooking an effective model in which
resonances are brought in and thrown away until a specific set fits
the data.  Several calculations, however, have shown that the data
cannot be reproduced without including some particular resonances,
thereby providing indirect support for the existence of these
excited states and their branching into the strange channels.

Apart from the choices with respect to the intermediate hadronic
states, an effective Lagrangian approach for the $p \left(\gamma, K^+
\right) \Lambda$ reaction involves the introduction of a set of 
coupling constants.  Being parameters in an effective theory, these
coupling constants can be calculated on the 
basis of QCD-inspired constituent-quark models for the hadrons
\cite{CapIsg,Capstick_2}. In this manner,  
the link between the $p \left( \gamma, K^+ \right) \Lambda$ data and
the quark models for baryons is established.  Accordingly, the
effective field theories allow to test 
theoretical predictions for coupling constants against
photo-production data.

Despite their success in reproducing the $p \left( \gamma, K^+ \right)
\Lambda$ observables over a photo-energy range from threshold up to
roughly 2~GeV, the hadronic models are facing a number of
difficulties. First, at photon lab energies above 2~GeV, the
predictions of all isobar models tend to overestimate the measured
cross sections. This feature is partly caused by $t$-channel
processes \cite{Guidal}.
Another difficulty is less well known and concerns the fact that the
Born terms in their own predict $p \left( \gamma, K^+ \right)
\Lambda$ cross sections which are  a few times the measured ones.  In
this paper we discuss various methods for counterbalancing the
strength produced by these Born terms. It turns out that hyperon
resonances can provide a natural mechanism to produce theoretical
cross sections  of the right order of magnitude.

The outline of this paper is as follows.  In Section
\ref{sec:isobar-model} the isobar model for $K^+$ photo-production on
the proton is briefly reviewed. In Section
\ref{sec:reaction-mechanism} the results of our numerical
$p(\gamma,K^+)\Lambda$ calculations are presented and the contribution
from the different terms in the reaction dynamics detailed.   Our
conclusions are presented in Section \ref{sec:concl}.

\section{The Isobar Model}
\label{sec:isobar-model}
\begin{figure}
\resizebox{0.5\textwidth}{!}{\includegraphics{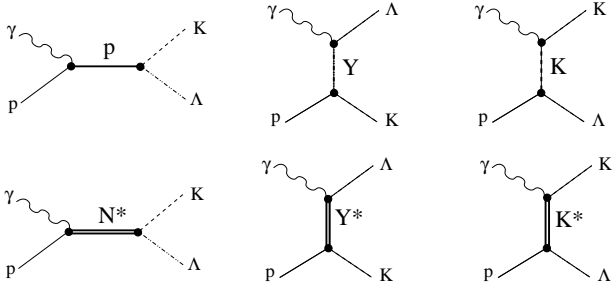}}
\caption{\em Diagrams contributing to the $p(\gamma,K^+)\Lambda$ process at
the tree level. The upper row corresponds to the Born terms in which a
proton is exchanged in the s-channel, a $\Lambda$ or $\Sigma^0$ in
the u-channel and a $K^+$ in the t-channel. The lower row shows 
the corresponding diagrams with the exchange of an excited particle or
resonance.} 
\label{fig:tree}
\end{figure}
Since the early work of Thom \cite{Thom} in the mid sixties, great
effort has been put into developing an isobar (or, hadronic) model for
the description of $p \left( \gamma, K^+ \right) \Lambda$
processes \cite{Adelseck,AdelSagh,Williams,David,Hsiao,Mart2}. Essentially, these 
effective field theories provide  
propagators for the intermediate particles and the structure of the
interaction Lagrangians describing the strong and electromagnetic vertices.
With this input one can compute the Feynman amplitudes ${\cal
M}^{\lambda_1 \lambda_2}_\lambda$ for a $ p(\lambda
_1) + \gamma (\lambda) \longrightarrow K^+ + \Lambda (\lambda _2)$ process
which determines the differential cross section (in the center-of-mass
frame) through the following
relation:
\begin{equation}
\frac{d \sigma}{ d \Omega} = \frac{1}{64
\pi^2} \frac{\left| \vec{p}_{\sss K} \right|}{ \omega} \frac{1}{W^2}
\frac{1}{4} \sum_{\lambda_1 \lambda_2
\lambda} \left| {\cal M}^{\lambda_1 \lambda_2}_\lambda \right|^2 \;,
\label{eq:cs_cm} 
\end{equation}
with $\lambda_1$, $\lambda_2$ and $\lambda$ respectively denoting the nucleon,
hyperon and photon polarization. Further, $W \equiv \sqrt{s}$ is the
invariant energy of the reaction.  At the tree level, the Feynman
amplitude is completely determined by the diagrams contained in
Fig.~\ref{fig:tree}. Here, we discriminate between the diagrams that
have hadrons in their ground-state ($p, \Lambda, \Sigma^0, K$) and those that
have hadron resonances ($N^*$, $\Lambda^*$, $\Sigma^*$, $K^*$) in the reaction
path.  In the isobar model, the 
composite nature of the hadrons is accounted for through the
introduction of form factors at every hadro\-nic vertex. A widely used
form for these hadronic form factors is \cite{Haberzettl,Pearce}:
\begin{equation}
F_x \left(\Lambda \right) = \frac{\Lambda^4}{\Lambda^4 + \left(x - M^2
\right)^2} \qquad (x \equiv s,t,u) \;,
\label{eq:formfac}
\end{equation}
where $x$ is the corresponding Mandelstam variable of the diagram in
question.  Further, $\Lambda$ is a cutoff parameter.  This cutoff
parameter sets a short-distance scale beyond which the hadronic model
is conceived to fail.  At best, the hadronic form factors may provide a
completely phenomenological description of the dynamical processes
which occur at 
distances smaller than those determined by the parameter $\Lambda$.  For the
sake of minimizing the number of free parameters, we introduce one
cutoff parameter for the hadronic vertices in the various Born terms
(diagrams in the upper row of Fig.~\ref{fig:tree}) and one for all
the resonant terms (diagrams of the lower row of
Fig.~\ref{fig:tree}).  To restore the broken gauge invariance after
introducing hadronic form factors, we
adopt\footnote{After the completion of the numerical calculations that
led to this work, Davidson and Workman \cite{Davidson} criticized some
aspects of the 
procedure of Haberzettl and suggested a different form factor for the
gauge breaking terms.}  a procedure suggested by Haberzettl {\em et al.}
\cite{Haberzettl}.

We have developed a computer program for the calculation of the
strangeness production observables.  With the aid of a symbolic trace
calculation, we first evaluate the expression for $ \left|
{\cal M}^{\lambda_1 \lambda_2}_\lambda \right| ^2 $ for the most general
case of pseudo-scalar meson photo-production.  Hereby, intermediate
vector mesons and baryons with $J=\frac{1}{2}$ and $\frac{3}{2}$ can be
accommodated. Second, the observables for $p
\left( \gamma, K^+ 
\right) \Lambda$ are computed numerically starting from the general
expression for $ \left| {\cal M}^{\lambda_1 \lambda_2}_\lambda \right|
^2 $ and specific choices with respect to intermediate particles,
coupling constants and cutoff masses.  A detailed outline of our model
will be presented in a forthcoming paper \cite{full_model}.

\section{Results and Discussion}
\label{sec:reaction-mechanism}

As mentioned in the introduction, at present it is not clear which
resonances represent the major contributions to the
$p(\gamma,K^+)\Lambda$ reaction mechanism.  From a kinematical point of
view, there are more than twenty likely candidates.  Over the years,
several combinations of resonances have been proposed in 
literature \cite{Adelseck,AdelSagh,Williams,David,Hsiao}. In all cases, good
agreement with the available data was achieved although the
conclusions and suggested sets of resonances were not unambiguous. This
diversity shows the complexity of the process and proves that after
more than three decades of research, there is still no established
reaction mechanism for the $p(\gamma,K^+)\Lambda$ process.

In our choice with respect to the intermediate particles, we have in
a first step been guided by recent coupled channel analyses
\cite{Mart2,Feuster2} that recognized the importance of three
intermediate states :  two spin 1/2 nucleon 
resonances ($N^*(1650)$ and $N^*(1710)$) and one spin 3/2 nucleon
resonance ($N^*(1720)$). Note that these nucleon resonances are also the
only ones in the particle-data tables \cite{PDG} with significant branching
into the strange channels. 
At higher energies \cite{Tran}, the
$p(\gamma,K^+)\Lambda$ data exhibit a typical diffractive nature. It
is well known that this feature is mainly due to $t$-channel
processes. For this reason, the two lowest vector meson resonances
($K^*(892)$ and $K_1(1270)$) are also explicitly included in our model
calculations. These five resonances ($N^*_{1650}$,
$N^*_{1710}$, $N^*_{1720}$, $K^*$ and $K_1$) constitute the basis of our
reaction dynamics. 
Nevertheless, an attempt to fit the cross sections and
polarization data with this basic set of five intermediate particles,
was only reasonably successful.  
Indeed, the agreement did not get any better than $\chi^2 \sim 10.32$
(see Table~\ref{tab:results}). This $\chi^2$ expresses the
conformity of the model calculations to the data.  
A value of 10.32 indicates that there is room for
other resonances as intermediate particles beyond the basic set of
five.

\begin{figure}
\resizebox{0.448\textwidth}{!}{\includegraphics{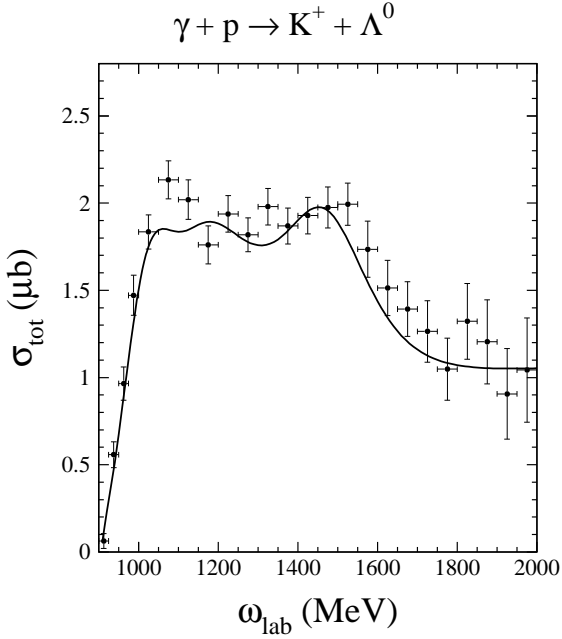}}
\caption{\em Our calculated result for the total cross
section in a model which includes 
the resonances $K^*$, $K_1$, $N^*_{1650}$, $N^*_{1710}$, $N^*_{1720}$
and $N^*_{1895}$ and uses a soft cutoff mass $\Lambda \sim 0.42$ GeV. The
data are from Ref.\cite{Tran}.}   
\label{fig:missres}
\end{figure}

The research in this field has experienced a new impulse with the advent
of the $p(\gamma, K^+)\Lambda$ data from the SAPHIR experiment at Bonn
\cite{Tran}. These data 
provide some indications for a structure in the $p\left( \gamma, K^+
\right) \Lambda$ total cross section about $\omega
_{lab}$ = 1500~MeV.  Due to limited energy resolution and statistics,
this structure could not be revealed in previous experiments. 
Recently, the Washington group \cite{Mart2} pointed out that model
calculations could account for this 
structure after including an additional spin 3/2 nucleon resonance
$N^*_{1895}$ in the $s$-channel.  The existence of this $D_{13}$
resonance with considerable branching into the strange channel, was
predicted by the constituent-quark model calculations of Capstick and Roberts
\cite{Capstick}.  Therefore, the authors of Ref.~\cite{Mart2}
legitimately claimed support for the existence of one of the
``missing resonances''.  Our calculations confirm the conclusions
drawn by the Washington group.  
When including the $N^*_{1895}$ in addition to the basic set of five
intermediate particles, we arrive 
at a promising $\chi^2 \sim 2.64$ (see Table~\ref{tab:results}).
The computed total cross section as a
function of the photon lab energy is given in Fig.~\ref{fig:missres}.

At this point, it is worth stressing that the quality of agreement
between the model calculations and the data very much depends on the
adopted value of the cutoff mass $\Lambda$ for the
Born terms.  As  
is mostly done, the cutoff mass is treated as a free parameter.
First, we considered the case where $\Lambda$ was allowed to vary
freely imposing an under limit of 0.4 GeV, though.  The best fit was reached
with a value $\Lambda \sim 0.42$ GeV, 
which corresponds to a soft form factor.  We also considered
the case where we forced the cutoff mass to be larger than 1.6~GeV,
which corresponds to an outspoken ``hard'' form factor. 
With the hard form factor, the overall agreement with the data was
inferior ($\chi ^2 \sim 7.38$) to 
what was achieved ($\chi ^2 \sim 2.64$) with the same set of
intermediate particles (basic 
set extended with $N^*_{1895}$) but with
a freely varying value of $\Lambda$.  From these observations, it is
clear that the hadronic form factors play a crucial role in the
reaction dynamics.  

\begin{figure}[b]
\resizebox{0.49\textwidth}{!}{\includegraphics{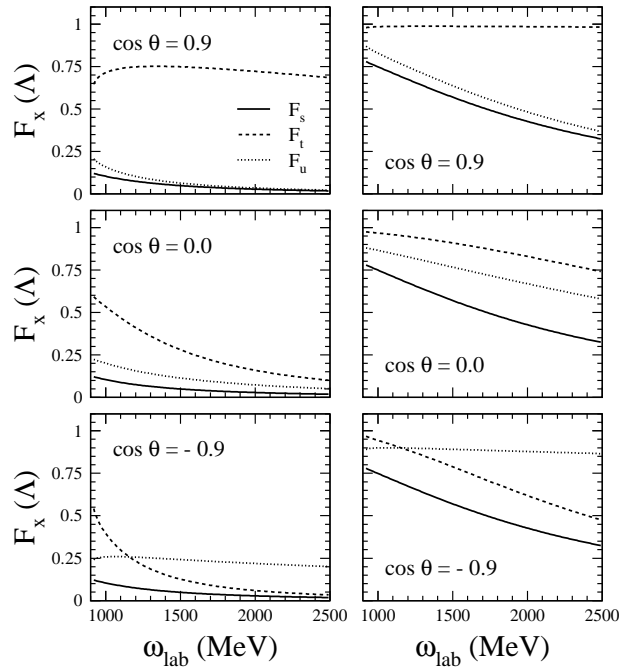}
}
\caption{\em The hadronic form factors $F_x(\Lambda)$ for two values
of the cutoff mass and for three kaon angles.  For the left panels
$\Lambda$ = 0.8 GeV, for the right panels $\Lambda$ = 1.8 GeV. The solid
line shows $F_s(\Lambda)$, the dashed line $F_t (\Lambda)$ and the
dotted line $F_u(\Lambda)$.}
\label{fig:ffdynam}
\end{figure}

We now attempt to figure out why the model with five or six
intermediate particles does a much  better job in reproducing the
data, when soft hadronic form factors are used.  To that purpose, we
separate the amplitude in so-called non-resonant parts represented
by the Born terms (upper row in Fig.~\ref{fig:tree}) and  resonant
parts, where resonances are the intermediate particles (lower row in
Fig.~\ref{fig:tree}). The non-resonant terms introduce two free 
parameters in the form of the coupling constants $g_{{\sst
\Lambda K} p}$ and $g_{{\sst \Sigma^0 K} p}$. Imposing SU(3) flavor symmetry,
these coupling constants can be related to the pion nucleon coupling
strength $g_{\pi {\sss N N}}$.  Despite the fact that SU(3) is a
broken symmetry, these relations can be used to impose a range of
values for the two coupling constants. Assuming that the symmetry
is broken at the level of $20\%$,  ranges are:
\begin{equation}
\begin{array}{rcccl}
-4.5\ &\leq &\ \ g_{\Lambda K p} / \sqrt{4 \pi}\ \  & \leq &\  -3.0 \;, \\
0.9\ & \leq &\ \ g_{\Sigma^0 K p} / \sqrt{4 \pi}\ \ & \leq &\ 1.3 \;.
\end{array}
\label{eq:boundcc}
\end{equation}
With values of $g_{{\sst \Lambda K} p}$ and $g_{{\sst \Sigma^0 K}
p}$ in these ranges, the non-resonant
terms are observed to produce an amount of strength which is at least
a factor of 4 or 5 larger than the measured cross sections. There are
a number of solutions to resolve this problem.  One possibility is to
let the coupling constants adopt values beyond the SU(3) limits. In
practice, this amounts to reducing $g_{{\sst \Lambda K} p}$ and
$g_{{\sst \Sigma^0 K} p}$ to (absolute) values far smaller than what
is expected within (broken) SU(3) \cite{Hsiao}.  Another possibility
is the introduction of a resonant 
term that interferes destructively with the non-resonant contribution.
A third possibility, adopted by the Washington group
\cite{Haberzettl}, is the introduction of a soft hadronic form factor
for the non-resonant terms.  The sensitivity of the hadronic form
factor to the value of $\Lambda$ is illustrated in
Fig.~\ref{fig:ffdynam} for a few values of the kaon center-of-mass
angle $\theta$.

As can be inferred from comparing the left and right panels in
Fig.~\ref{fig:ffdynam}, the introduction of soft hadronic form factors
induces a severe reduction of the strength attributed to the Born
terms. As a consequence, when the cutoff mass is chosen sufficiently
low, the coupling constants can be kept between their SU(3) limits and
there is no need for an extra resonant term to 
counterbalance the strength from the Born diagrams.  It
appears, however, that the cutoff masses at which the strength from
the non-resonant terms can be sufficiently suppressed, are of the order
of the kaon mass. These small values of $\Lambda$ may raise 
some questions with respect to the realistic character and
applicability of the ``effective'' theory.  Indeed, as mentioned
above, the value of the cutoff mass sets a short-distance scale to the
theory and we believe that a $\Lambda = 0.8$ GeV is probably already
an under-limit of what appears to be a reasonable cutoff
mass. Moreover, the severe reduction of the Born strength through the
introduction of a soft form factor (see
Fig.~\ref{fig:ffdynam}) may appear as a rather artificial way of keeping
$g_{{\sst \Lambda K} p}$ and $g_{{\sst \Sigma^0 K} p}$ in agreement
with the SU(3) expectations.  

A value of $\Lambda \sim m_{\sss K}$ also appears to be rather small
in comparison with  
the typical values which are predicted in potential models for the
hadron forces. 
For example, the values for $\Lambda$ quoted by the Nijmegen
\cite{Rijken} and the J\"ulich group \cite{Holznkamp} are close to 1.2
GeV. Recently the J\"ulich group \cite{Haidenbauer} even proposed  cutoff
masses in the meson-baryon sector exceeding 2 GeV. 

\begin{figure*}
\begin{center}
\resizebox{0.95\textwidth}{!}{\includegraphics{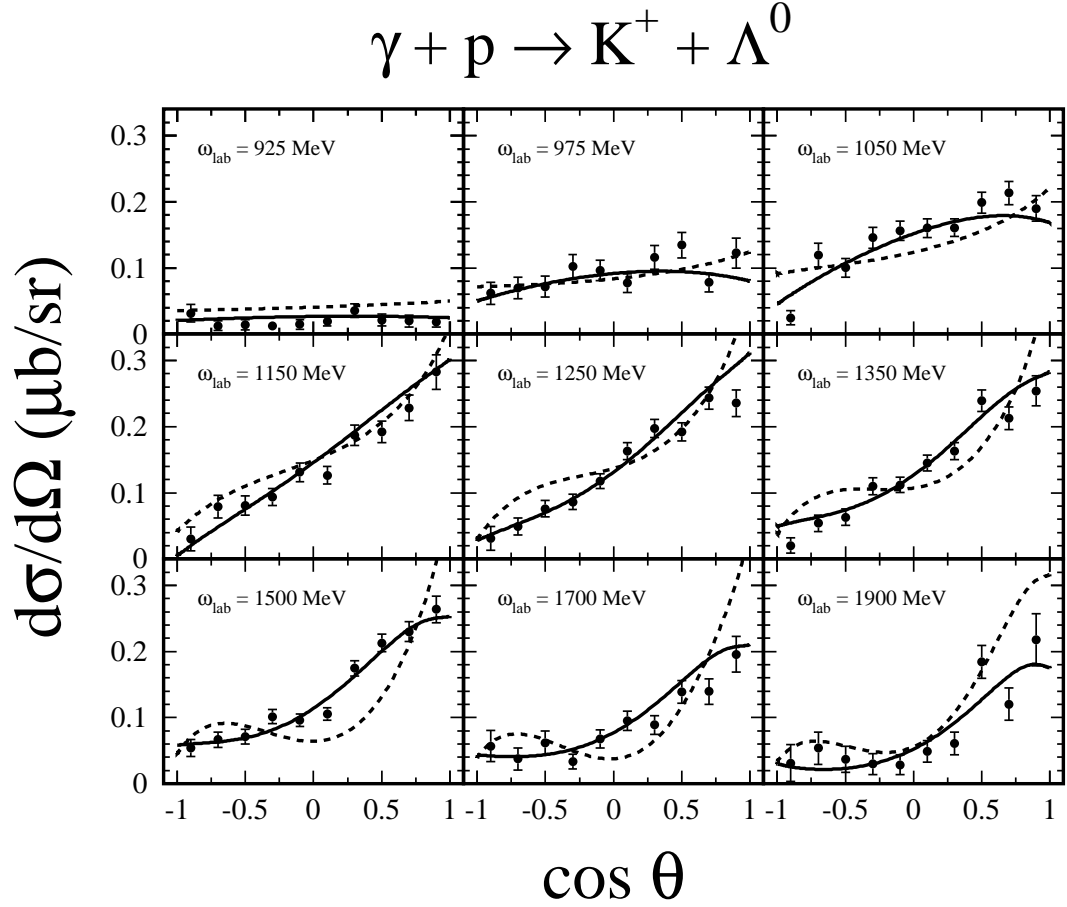}}
\caption{\em Model calculations for the differential cross section at
various photon energies. The dashed curves is a
calculation with the ``basic set + $N^*_{1895}$''. For the solid curves
this set was extended with two hyperon resonances ($\Lambda^*_{1800}$
and $\Lambda^*_{1810}$). A hard form factor ($\Lambda \geq$ 1.6 GeV) is
used in both model calculations. The data are from the SAPHIR collaboration 
\cite{Tran}.} 
\label{fig:diffcs}
\end{center}
\end{figure*}
\begin{figure*}
\begin{center}
\resizebox{0.95\textwidth}{!}{\includegraphics{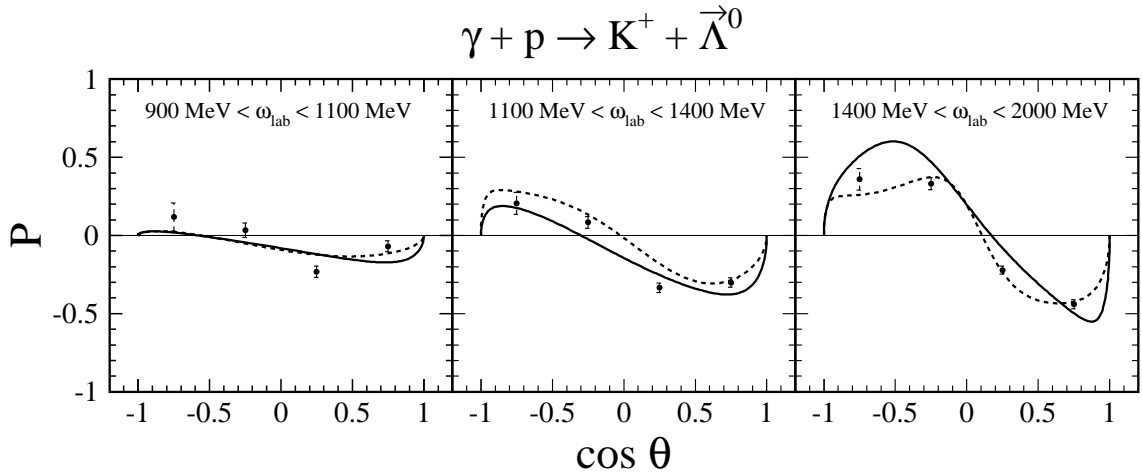}}
\caption{\em Model calculations for the recoil polarization
asymmetry at various photon energies. The curves are
as in Fig.~\ref{fig:diffcs}. The data are from the SAPHIR
collaboration \cite{Tran}.} 
\label{fig:recpol}
\end{center}
\end{figure*}

Here, we suggest an alternative method to counterbalance the
unrealistically large amounts of strength produced by the non-resonant
Born diagrams.  We find that, after the
inclusion of the two spin 1/2 hyperon resonances
$\Lambda^* (1800)$ and $\Lambda^* (1810)$ (both received a
$(***)$-ranking by the Particle Data Group \cite{PDG}) in the 
$u$-channel, the overall $\chi^2$ for the complete data set drops 
from 7.38 to 2.65 (see Table~\ref{tab:results}). This agreement with
the data is as favorable as the one obtained with the ``basic set +
$N^*_{1895}$'' in combination with a soft hadronic form factor. 
We observe that after including the hyperon resonances, good fits to
the data were achieved with hard form factors corresponding 
with typical values of $\Lambda \sim$ 1.8 GeV. 
The coupling strengths of these two intermediate $\Lambda^*$ states
turns out to be relatively large in comparison with the typical values
obtained for the nucleon resonances couplings.
We have also tried other combinations 
of known spin 1/2 hyperon resonances in the $u$-channel. All
combinations improved the global agreement between the calculations
and the complete data set. Furthermore, in all cases a similar qualitative
interference pattern between the other terms was observed. The
combination of $\Lambda^*_{1800}$ and $\Lambda^*_{1810}$, though,
produced the best $\chi^2$.
\textit{Therefore, the suggested procedure
of including hyperon resonances in the $u$-channel emerges as a natural
way of compensating for the strength produced by the Born terms
and allows the cutoff masses to have more realistic values.}

\begin{figure}
\resizebox{0.5\textwidth}{!}{\includegraphics{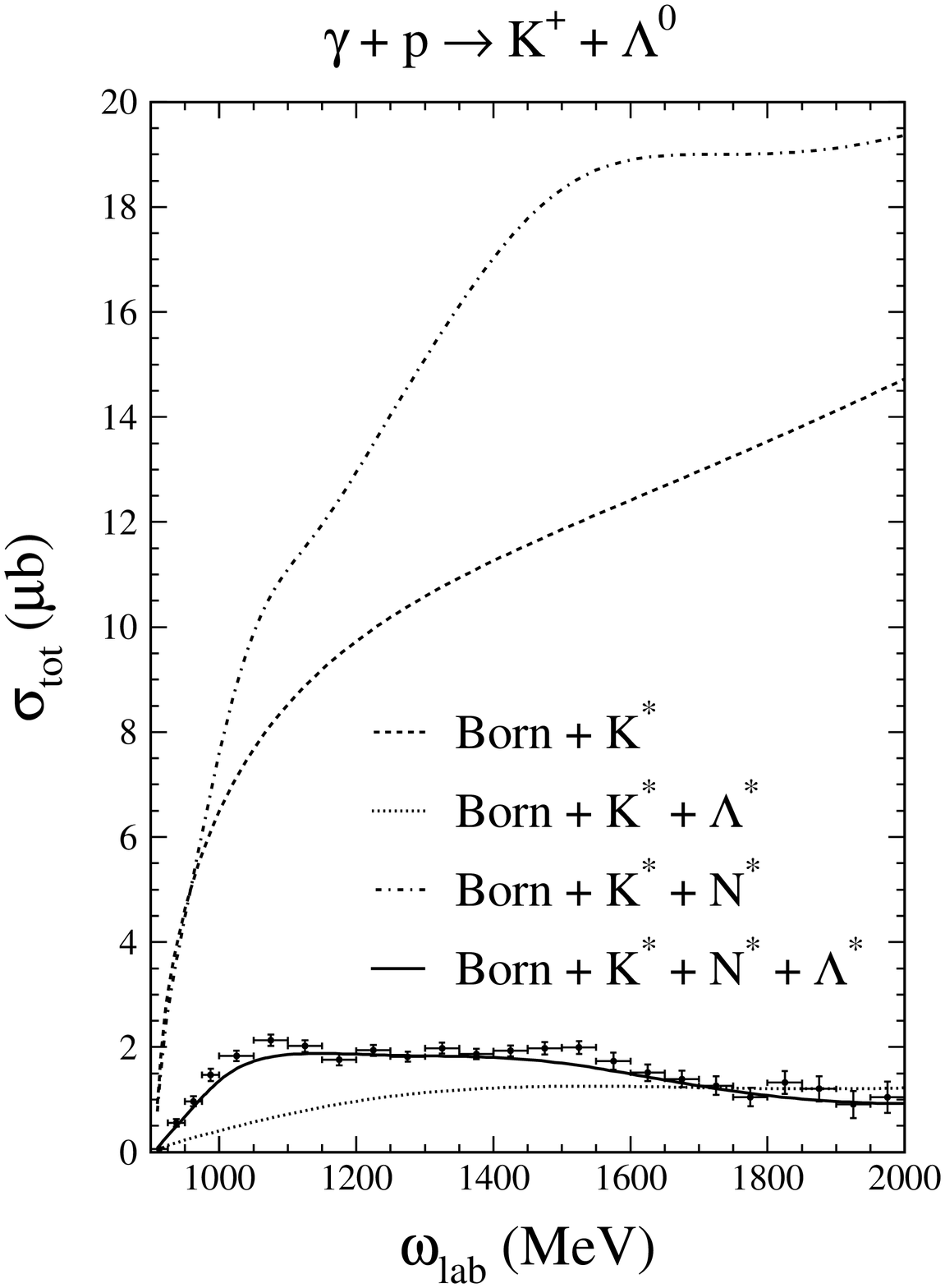}}
\caption{\em The total $p(\gamma,K^+)\Lambda$ cross section as a function of
the photon lab energy. The dashed line gives the combined Born and
vector meson strength, the dot-dashed line includes in addition nucleon
resonances in the s-channel. The dotted curve shows the effect of
including  hyperon resonances to the Born and vector meson
background. The solid line is a calculation with the complete
set. The cutoff mass in these model calculations was $\Lambda \sim 1.78$ GeV.}
\label{fig:hypres}
\end{figure}

As illustrated in
Fig.~\ref{fig:diffcs} and Fig.~\ref{fig:recpol}, after inclusion of the
hyperon resonances in the reaction dynamics a fair agreement between
the calculations and the 
data is reached ($\chi^2 \sim 2.65$). Admittedly, although the
$N^*_{1895}$ is explicitly included, the
calculations (solid line in Fig.~\ref{fig:hypres}) do not longer
predict an outspoken structure in the total 
cross section about $\omega _{lab}$ = 1500~MeV.

The origin of the successes achieved with the introduction of hyperon
resonances can be traced back to a strong destructive interference
between the contribution from the hyperon resonances and the Born
terms. This can be inferred from a detailed examination of
Fig.~\ref{fig:hypres} which shows the contributions to
the total cross section from the
different types of resonances (vector meson resonances K$^* \equiv
(K^*,K_1)$, nucleon resonances N$^* \equiv (N^*_{1650}$,$N^*_{1710}$,
$N^*_{1720}$, $N^*_{1895})$,  and hyperon resonances $\Lambda^* \equiv
(\Lambda^*_{1800}$, $\Lambda^*_{1810})$). 
It should be stressed that the different curves do not represent the
best fit to the data for a particular combination of intermediate
particles. The curves show the predicted strength from a specific
combination of diagrams, when fixing the parameters with values that
are obtained in a calculation that includes the ``basic set + $N^*_{1895}$,
$\Lambda^*_{1800}$, $\Lambda^*_{1810}$''. So, the different curves in
Fig.~\ref{fig:hypres} illustrate how the final result (solid curve)
comes about as a coherent sum of several contributing diagrams.
The combination of Born terms (with coupling
constants constrained within the SU(3) ranges of
Eq.~(\ref{eq:boundcc})) and vector mesons produces far too 
much strength in comparison with the measured cross sections. 
After including the hyperon resonances
(denoted by ``Born+K$^*$+$\Lambda ^*$''), the computed
cross section has already the right order of magnitude due to a
destructive interference.  Further
inclusion of the nucleon resonances in the $s$-channel produces the
required structure of the cross section but does not sizably alter
the magnitudes. 
Including all terms but the $\Lambda^*$'s in the $u$-channel (a
calculation denoted by ``Born+K$^*$+N$^*$'') we obtain cross sections
that overshoot the data by almost an order of magnitude.
Concluding, it is essential to realize that hyperons in the
$u$-channel are likely candidates for playing a predominant role in
the $p(\gamma,K^+)\Lambda$ reaction dynamics. 
\begin{table}[b]
\begin{center}
\begin{tabular}{lcc}
\hline \hline
& & \\
Resonance set           & $\Lambda$ (GeV)  & $\chi^2$ \\ 
& & \\
\hline
& & \\
basic set                   &  $\geq$ 1.6  &  10.32   \\ 
basic set                   &  $\geq$ 0.4  &  4.36  \\
basic set $+\ N^*_{\sss 1895}$ & $\geq$ 1.6 &  7.38   \\ 
basic set $+\ N^*_{\sss 1895}$ & $\geq$ 0.4  &  2.64   \\ 
basic set  $+\ \Lambda^*_{\sss 1800}$ $\Lambda^*_{\sss 1810}$ &$\geq$1.6 
& 3.43 \\  
basic set $+\ N^*_{\sss 1895}$ $\Lambda^*_{\sss 1800}$ $\Lambda^*_{\sss 1810}$
& $\geq$ 1.6 &  2.65   \\ 
& & \\
\hline \hline
 & &  
\end{tabular}
\caption{{\em Overall agreement between the model calculations and the
SAPHIR data for the different sets of resonances and cutoff masses.
We denote by ``basic set'' the following set of intermediate particles :
$K^*, K_1, N^*_{\sss 1650}, N^*_{\sss 1710}, N^*_{\sss 1720}$}.}
\label{tab:results}
\end{center}
\end{table}

We summarize our findings in Table~\ref{tab:results} where we give an
overview of the best $\chi^2$ values which we achieved for the various
combinations of resonances. For these calculations we have used all the
SAPHIR data, including the complete set of total- and differential
cross sections and recoil 
polarization asymmetries. Without the introduction of a hyperon
resonance in the reaction dynamics, a soft hadronic form factor
produces a far better description of the data than a hard one.
A $\chi ^2$ calculation with the 
basic set of five intermediate particles ($K^*, K_1, N^*_{\sss
1650}, N^*_{\sss 1710}, N^*_{\sss 1720}$) improves from 10.32 to 4.36
when a soft instead of a hard
hadronic form factor is used. In both cases (hard and soft cutoff
masses) the $\chi 
^2$ further decreases after 
adding the ``missing'' $N^*_{1895}$ resonance to the basic set
($\chi ^2$ = 7.38 for $\Lambda \geq 1.6$ GeV, $\chi
^2$ = 2.64 for $\Lambda \geq 0.4$ GeV).  One way of obtaining
reasonable fits with hard hadronic form factors is allowing
hyperon resonances in the reaction dynamics. 

Inspecting Table~\ref{tab:results}, one observes that after implementing
hyperon resonances in the reaction dynamics, the supplementary
introduction of the $N^*_{1895}$ does only lead to a minor improvement
in the quality of the description of the data, despite the fact that 5
additional  parameters are introduced in the fitting procedure.

\begin{figure*}
\resizebox{1\textwidth}{!}{\includegraphics{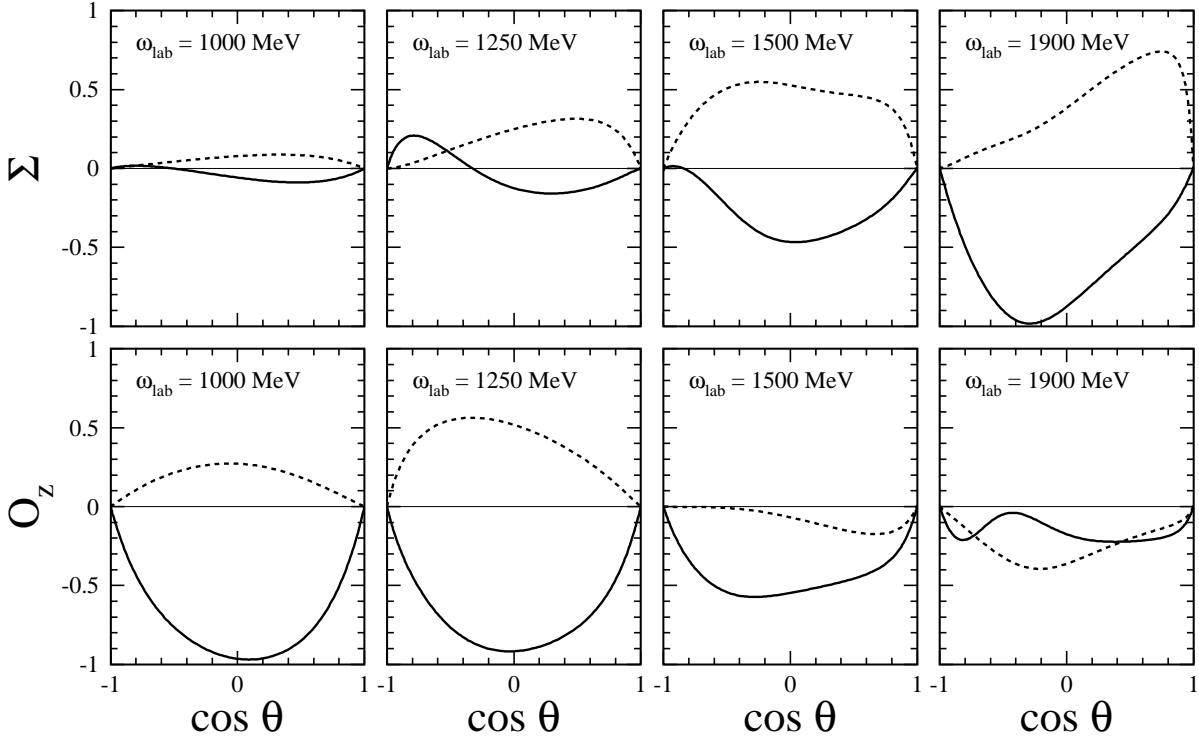}}
\caption{\em Model predictions for the photon polarization asymmetry
$\Sigma$ and the double polarization beam-recoil asymmetry $O_z$ at
various photon lab energies.  The dashed curves are calculations for the
``basic set + $N^*_{1895}$'' and a cutoff mass
$\Lambda \sim$ 0.42 GeV. The solid curves are calculations where the
same set is extended with $\Lambda^*_{1800}$ and $\Lambda^*_{1810}$
and a cutoff mass  $\Lambda \sim$ 1.78 GeV.}
\label{fig:comb_asym}
\end{figure*}

From the above discussion it becomes clear that as far as it comes to
reproduce 
the differential cross sections, the two approaches (soft hadronic
form factors or hyperon resonances and hard hadronic form factors)
produce comparable results.  This is, 
however, not the case for all the observables.  To illustrate this, we
present model calculations for single and double polarization
asymmetries. Fig.~\ref{fig:comb_asym} shows predictions for the
photon polarization asymmetry $\Sigma$ and the double beam-recoil
asymmetry $O_z$. 
These quantities are defined as:
\begin{eqnarray}
\Sigma &=& \frac{ d \sigma / d \Omega^{\left( \perp \right)} - d
\sigma / d \Omega^{\left( \parallel \right)} }{ d \sigma / d
\Omega^{\left( \perp \right)} + d \sigma / d \Omega^{\left( \parallel
\right)} } \;, \label{eq:sigma}
\\
O_z &=& \frac{ d \sigma / d \Omega^{\left(++ \right)} - d \sigma / d
\Omega^{\left(+- \right)}}{ d \sigma / d \Omega^{\left(++ 
\right)} + d \sigma / d \Omega^{\left(+- \right)}} \;.
\label{eq:oz}
\end{eqnarray}
In Eq.~(\ref{eq:sigma})  $\perp \left(\parallel  \right)$ refers 
to linearly polarized photons perpendicular (parallel) to the reaction
plane. In Eq.~(\ref{eq:oz}) the first $+ (-)$ corresponds with linearly
polarized photons under an angle of $ + (-) \pi/4$ with respect to
the scattering plane. The second $+ (-)$ refers to recoil
polarization parallel (anti parallel) to the momentum of the escaping
$\Lambda$.
In different energy regions, both observables seem to be
extremely sensitive to the introduction of hyperon resonances and the
magnitude of the 
adopted cutoff masses.

\section{Conclusions}
\label{sec:concl}
We have suggested an alternative technique to counterbalance the
unreasonably high 
strength which is produced by the Born terms in effective-field
approaches to the $\gamma + p \rightarrow K^+ + \Lambda$ process.  It
involves both the introduction of hyperon resonances in the $u$-channel
and form factors at the hadronic vertices.  Herewith, we obtain a fair
description of the SAPHIR cross sections and polarization observables
with values of the hadronic cutoff mass that are a few times the kaon mass.
Alternative approaches to counterbalance the strength from the Born terms,
either go out from soft hadronic form factors with cutoff masses
that are of the order of the kaon mass or use coupling constants
smaller than those predicted by SU(3) constraints.
We have observed that the presence of the hyperon
resonances in the $u$-channel has a large impact on the predicted
values for the polarization
observables. 
We believe that the
measurement of these quantities could distinguish between the
different models for the underlying dynamics of the strangeness
production process and could shed light on the value of the cutoff masses.

\begin{acknowledgement}
This work was supported by the Fund for Scientific Research - Flanders
under contract number 4.0061.99. 
\end{acknowledgement}

\bibliographystyle{osa}

\end{document}